\documentclass[a4paper,10pt]{article}
\pdfoutput=1 

\usepackage{jheppub} 
\usepackage{amssymb,amstext,amsthm,amsfonts,mathrsfs,amsbsy,amsmath,array,multirow}
\usepackage[T1]{fontenc} 

\usepackage{graphicx}

\preprint{IPhT-t15/210}

\title{\boldmath Five-dimensional null \& time-like supersymmetric geometries}

\author{Giulio Pasini and C. S. Shahbazi}

\affiliation{Institut de physique th\'eorique, Universit\'e Paris Saclay, CEA, CNRS, F-91191 Gif-sur-Yvette, France}

\emailAdd{giulio.pasini@cea.fr, carlos.shabazi-alonso@cea.fr}

\abstract{We show that there exist supersymmetric solutions of five-dimensional, pure, $\mathcal{N}=1$ Supergravity such that the norm of the supersymmetric Killing vector, built out of the Killing spinor, is a real not-everywhere analytic function such that all its derivatives vanish at a point where the Killing vector field becomes null. The norm of the Killing vector field then is not an analytic function on a neighborhood around this point. We explicitly construct such solutions by using a multi-center Gibbons-Hawking base. Although many of these solutions have infinite charges, we find explicit examples with finite charges that asymptote to $AdS_3\times S_2$ and discuss their physical interpretation.}

\begin{document}
\maketitle
\flushbottom


\section{Introduction}


Supersymmetric solutions of minimal Supergravity in five dimensions play an essential role in various areas of String Theory. For instance, they are fundamental to understand String Theory compactifications as well as the microscopic properties of black hole solutions. In addition, they are usually good toy models to understand the key properties and features of more complicated higher dimensional solutions. In fact, five-dimensional, pure, $\mathcal{N}=1$ Supergravity is the perfect framework where to explore the geometry and topology of M-theory supersymmetric solutions~\cite{Mizoguchi:1998wv}. 

The supersymmetric solutions of five-dimensional, pure, $\mathcal{N}=1$  Supergravity were classified in~\cite{Gauntlett:2002nw} up to local isometries, while the local classification in the complete, matter-coupled, five-dimensional Supergravity was found in~\cite{Gauntlett:2003fk,Bellorin:2006yr,Bellorin:2007yp}. Locally, these solutions can be divided into two classes according to the causal character of the supersymmetric Killing vector field, namely the vector field given as a bilinear in terms of the Killing spinor\footnote{To save unnecessary words, we will refer to the supersymmetric Killing vector as simply the Killing vector.}. The \textit{time-like class} is characterized by a time-like Killing vector, while the \textit{null class} is characterized by a null Killing vector\footnote{It can be shown that the Killing vector cannot be spacelike.}. In the time-like class, supersymmetry constrains the metric in such a way that it can be formally written in terms of a local Hyper-K\"ahler four-dimensional \emph{base space}.

It is well-known that in the time-like class the Killing vector can become null at some loci of the four-dimensional Hyper-K\"{a}hler base manifold. Indeed, this is precisely what happens at the horizon of a black hole solution, or in the smooth five-dimensional solutions of~\cite{Bena:species} which will be considered in this letter. A change in the causal character of the Killing vector is paralleled by a change of the supersymmetry conditions that the spinor satisfies. However, in all the solutions constructed so far this can happen only in some regions of codimension at most one of the space-time manifold, that are typically surfaces. 

Our goal in this letter is to construct five-dimensional solutions in minimal Supergravity whose Killing vector, which is generically time-like, becomes null at a point of the space-time manifold where all the derivatives of its norm vanish. The null condition is a closed condition, so if the norm of the spinor is a continuous function then the spinor can become null only on a closed subset of the manifold\footnote{We thank Jos'e Figueroa-O'Farrill for clarifications about this point.}. Since all the derivatives of the norm vanish at the point where the Killing vector field becomes null, we conclude that the norm is not an analytic real function at this point, otherwise the Killing vector field would vanish on an open set, contradicting the fact that the null condition on the norm is a closed condition. Therefore, if we are able to construct a solution where the norm of the Killing vector field and all its derivatives vanish at some locus, we will know that the norm is not a real analytic function at those points. In either case, having a Killing vector whose norm has an infinite number of derivatives vanishing at a point is the closest scenario to having a null-spinor on an open set. These would be solutions that may mix, in a non-trivial way, the local classification that distinguishes between the time-like and the null classes. In this respect we consider smooth Supergravity solutions with a multi-center Gibbons-Hawking (GH) base manifold that asymptote to $AdS_3\times S_2$. These can be generated from the compactification of eleven-dimensional Supergravity solutions on three tori with stabilized moduli. The conditions that ensure smoothness were found in~\cite{Bena:bubbling_supertubes, Berglund:microstates} as these solutions represent microstate geometries for five-dimensional black holes. These solutions are uniquely determined once one fixes the poles and the residues of two harmonic functions $V$ and $K$ in the GH space. In particular, it is well known~\cite{Bena:species} that the related Killing vector is time-like almost everywhere, except for the loci where $V$ is zero. 

It is easy to give a physical interpretation for this phenomenon in the eleven-dimensional formalism: in the regions where the Killing vector is time-like the supersymmetry conditions are compatible with those of M2 branes wrapping one of the three tori. At the same time, on the surfaces where the Killing vector becomes null, the supersymmetry conditions become those required by M5 branes wrapping two of the three tori and the Gibbons-Hawking fiber. The transition between M2 brane-like supersymmetry and the M5-brane one can be of utmost importance for the construction of new microstate geometries for five-dimensional black holes that are not electrically charged. This can be achieved by means of a new class of physical objects, the magnetubes, introduced in~\cite{Bena:species}. A magnetube has an M5 charge, which is magnetic in five dimensions, together with positive and negative M2 charges. The M2 charge density is allowed to smoothly vary along the M5-P common direction so that the net M2 charge is zero. The supersymmetry conditions are those for M5 branes along this direction and hence the positive and negative M2 charges do not interact and the whole magnetube is supersymmetric. This can lead to the construction of microstates for five-dimensional black holes with zero electric charges~\cite{Mathur:2013nja}, such as the Schwarzschild ones. Finding a solution where the Killing vector becomes null at a point where its norm has an infinite number of vanishing derivatives can be relevant in this respect. Indeed, this solution would allow to construct new types of magnetubes, as the M5 brane charge is no longer constrained in the standard way. In addition, one could add different types of magnetubes in the same null region and study whether the counting of these types of configuration can  partially reproduce the expectations from the black hole entropy.

Another motivation for studying global properties of five-dimensional Supergravity solutions comes from the analysis on the global geometry and topology of $AdS_{3}\times M_{8}$ supersymmetric solutions of eleven-dimensional Supergravity~\cite{Babalic:2014fua,Babalic:2015kka,Babalic:2014dea,Babalic:2015xia}. Indeed, the structure of this class of supersymmetric solutions is extremely rich and heavily depends on the properties of the Killing spinor field on the space-time manifold. Since the study of the topology and geometry of this class of solutions is very involved, it is natural to first explore their five-dimensional counterparts as toy models for the eleven-dimensional scenario.

To reach our goal it is hence necessary to construct a five-dimensional solution that asymptotes to $AdS_3\times S_2$ where the harmonic function $V$ vanishes at a point together with all its derivatives. This operation is in general impossible, unless one allows the number of poles to become infinite. We then adopt the following strategy. We arrange $2N$ poles on the same line in the GH space so that the function $V$ and all its derivatives up to order $2N-2$ vanish at the origin of the GH space. This constrains the residues of $V$ at the poles to be determined by the distances $d_i$'s between the poles and the origin. In the limit where $N$ becomes infinite $V$ and all its derivatives vanish at the origin, where the Killing vector becomes null. 

The choice of the distances $d_i$'s of the poles from the origin is the only degree of freedom left by our construction and the physical relevance of the solution given by the limit heavily depends on this. Indeed, a general requirement to determine whether the solution in the limit is physically meaningful is to demand it to asymptote to $AdS_3 \times S_2$, so that it still belongs to the original class of five-dimensional solutions. We give numerical evidence that there exists a distance distribution for the poles so that this condition is satisfied. At the same time, the residues of $V$ remain finite in the limit, while all the poles collapse on two different points. This result motivates a future analysis about the behavior of the metric and the warp factor in the proximity of these two would-be essential singularities, that in fact might not be singularities at all. Indeed, considering the full backreaction what appears to be an essential singularity from the point of view of the three-dimensional base of the GH space will in fact give rise to smooth solutions~\cite{Bena:mergers, Bena:abyss}.

This paper is organized as follows. In Section~\ref{sec:5dSugra} we present a general review about the supersymmetric structure of $\mathcal{N}=1$ five-dimensional Supergravity solutions, with emphasis on the distinction between time-like and null classes. In Section~\ref{sec:model} we recall the construction of smooth five-dimensional solutions with a Gibbons-Hawking base that asymptote to $AdS_3 \times S_2$ and prove that the associated Killing vector becomes null when $V$ vanishes. In Section~\ref{sec:construction} we show that by suitably arranging $2N$ GH centers on a line it is possible to have $V$ as well as all its derivatives up to order $2N-2$ vanish at the origin. In Section~\ref{sec:limit} we consider the limit where $N$ becomes infinite, so that all the derivatives of the norm of the Killing vector vanish at the origin. In particular, we numerically show that it is possible to arrange the distances between the GH centers so that the limit solution still asymptotes to $AdS_3\times S_2$. In Section~\ref{sec:conclusions} we underline the physical properties of the limit solution and state some future work. Additional details are presented in the Appendix. In particular, in Appendix~\ref{appendix:convergence} we prove that the residues of $V$ attain a finite value in the limit, while in Appendix~\ref{appendix:external} we briefly describe an alternative construction that does not lead to a physically relevant limit solution.


\section{The supersymmetric solutions of $\mathcal{N}=1$ five-dimensional Supergravity}
\label{sec:5dSugra}


In this section we review the structure of the supersymmetric solutions of five-dimensional $\mathcal{N}=1$ pure Supergravity. Although the theory under scrutiny here seems to be relatively simple, the structure of its supersymmetric solutions is remarkably rich. This fact can be traced back to the particular form of the Killing spinor equation, which is relatively involved\footnote{In particular, it is not the \emph{lift} to the spin bundle of a metric connection.} but also to the quaternionic structure of the spinor bundle of the solutions.

We will focus exclusively on bosonic solutions. The bosonic matter content of five-dimensional, pure, $\mathcal{N}=1$ Supergravity consists of a Lorentzian, oriented, spin manifold $(M_{5},g_{5})$ together with a connection $A$ on principal $U(1)$-bundle $P\to M_{5}$ over $M_{5}$. The bosonic part of the action is given by:

\begin{equation}
\mathcal{S} = \int_{M_{5}} \left\{ R - \frac{1}{4} |F|^2 + \frac{1}{12\sqrt{3}} F\wedge F\wedge A\right\} 
\end{equation}

\noindent
The equations of motion of the theory are given by:

\begin{eqnarray}
\label{eq:eqs5}
g(u,v) + \frac{1}{2}\left( < F(u), F(v)> -\frac{1}{4} g(u,v) < F, F>\right) &=& 0    \qquad u,v \in \mathfrak{X}(M_{5})\nonumber \\
d\ast F + \frac{1}{4\sqrt{3}}\, F\wedge F &=& 0   \qquad \alpha\in\mathbb{R}^{\ast}
\end{eqnarray}

\noindent
where $F$ is the curvature associated to $A$ and $< \cdot, \cdot>$ denotes the inner product on forms induced by $g$. A pair $(g_{5}, A)$ satisfying equations~\eqref{eq:eqs5} is said to be a bosonic solution of the theory. 

Let $Cl(M_5,g_5)$ denote the bundle of real Clifford algebras over $(M_5,g_5)$, isomorphic as a bundle of, unital, associative algebras to the K\"ahler-Atiyah bundle $(\Lambda T^{\ast}M_5, \diamond)$, see references~\cite{Lazaroiu:2012du, Lazaroiu:2012fw, Lazaroiu:2013kja} for a detailed account of this formalism. Let us assume in addition that there exists a bundle of real Clifford modules $S$ over $M_5$ with representation homomorphism denoted by:

\begin{equation}
\gamma\colon (\Lambda T^{\ast}M_5, \diamond)\to (End(S),\circ)
\end{equation}

\noindent
where $(End(S), \circ)$ denotes the unital, associative, algebra of endormophisms of $S$. In Lorentzian signature in five dimensions, $\gamma$ is neither surjective nor injective~\cite{Lazaroiu:2013kja}, and $S$ is a rank-eight bundle of real Clifford modules, which remains irreducible as a spinor bundle for $Spin(1,4)$. The commutant subbundle $\Sigma$ of the K\"ahler-Atiyah bundle inside the endomorphisms of $S$ is of quaternionic type. This implies that on every open set $U\subset M_{5}$ there exists a local triplet $J^{i}, i = 1, 2, 3,$ of almost complex structures satsifying the algebra of the imaginary quaternions. Notice however that these almost-complex structures do not exist globally and thus $\Sigma_{\gamma}$ is in general not topologically trivial. This is already a hints that the supersymmetric solutions of the theory may have very subtle non-trivial properties at the global level.

Since $Cl(M_5,g_5)$ is non-simple, there are two inequivalent Clifford modules $S$, distinguished by the value of the volume element $\gamma(\nu)$ inside $End(S)$:

\begin{equation}
\gamma (\nu) = s_{\gamma}\, Id   \qquad s_{\gamma}\in \left\{ 1, -1 \right\}
\end{equation} 

\noindent
We will take the $s_{\gamma} = 1$ in the following. The bundle of Clifford modules $S$ can be endowed with a symmetric admissible bilinear $\mathcal{B}$, that is of utmost importance in order to write a spinor as a polyform. A bosonic solution $(g_{5}, A)$ is said to be supersymmetric if there exists a globally defined spinor\footnote{Strictly speaking, this is a pinor, namely a module for the Clifford algebra. This fact may have important consequences, since the topological obstruction for the existence of pinors is weaker than the topological obstruction for the existence of spinors. In particular, the manifold may no be orientable and still admit pinors.} $\epsilon\in \Gamma(S)$ satisfying:

\begin{equation}
\label{eq:KSE5}
\nabla_{u} \epsilon - \frac{1}{8\sqrt{3}}\, u^{\flat}\wedge F\cdot\epsilon + \frac{1}{2\sqrt{3}}\, \iota_{u}\, F\cdot \epsilon = 0  \qquad \forall u\in\mathfrak{X}(M_{5})
\end{equation}

\noindent
A pair $(g_{5},A)$ for which there exists a globally defined spinor $\epsilon\in \Gamma(S)$ satisfying \eqref{eq:KSE5} is said to be a \emph{supersymmetric configuration}. Using the results of references \cite{Lazaroiu:2013kja,Reconstruction}, we conclude that a pinor in five Lorentzian dimensions is equivalent a polyform:

\begin{equation}
E \in \Omega^{\bullet}(M_5)
\end{equation}

\noindent
satisfying the \emph{generalized Fierz relations}\footnote{These relations are not always equivalent to the standard Fierz relations appearing in the physics literature. See reference \cite{Pinfoliations} for more details.}. The polyform $E$ can be written in terms of a function $Z$, the supersymmetric Killing vector $p$ and a triplet of two-forms $\Phi^{s}$ depending only on the coordinates of the base space. In turn, these can be locally written in terms of the admissible bilinear form $\mathcal{B}$ and the local triplet of almost-complex structures $J^{i}$, local sections of $\Sigma$ which exist precisely because $\Sigma$ is of quaternionic type. The local expressions of $Z$, $p$ and $\Phi^{s}$ in terms of a local frame are:

\begin{align}
Z^{-1}& = \mathcal{B}_{0}(\epsilon , \epsilon)    &  p_{a} &= \mathcal{B}_{0}(\epsilon , \gamma_{a}\epsilon)    &      \Phi^{i}_{ab} &= \mathcal{B}(\epsilon, J^{i}\circ \gamma_{ab} \epsilon) \label{bilinear}
\end{align}

\noindent
The Killing spinor equation~\eqref{eq:KSE5} translates into a set of differential conditions for $Z$, $p$ and $\Phi^{i}$. In reference \cite{Gauntlett:2002nw} the most general local form of a supersymmetric solution of five-dimensional pure, $\mathcal{N}=1$ Supergravity was obtained\footnote{See also reference \cite{Bellorin:2006yr} for the most general local form of the supersymmetric solutions of $\mathcal{N}=1$ five-dimensional Supergravity coupled to vectors and hypers.}. The supersymmetric solutions can be divided in two classes: the \textit{time-like class} is characterized by a time-like Killing vector $p$, while the \textit{null class} is characterized by a null Killing vector. Notice that this classification is local and that the two classes can overlap. From the Fierz algebra one gets:

\begin{equation} 
\label{Fiertz}
g_{5}(p,p) = - Z^{-2}
\end{equation}

\noindent
and hence $p$ cannot be space-like. The local form of the solution in each class is as follows:

\begin{itemize}
\item {\bf Null class}. There exist local coordinates $u, v, x^{s}$ with $s=1,2,3$ such that the metric can be written as:

\begin{equation} \label{null_class}
ds^{2} = -Z^{-1} du (dv + H\, du + \omega) + Z^{2} \delta_{rs} dx^{r} dx^{s}
\end{equation}

\noindent
where $Z$, $H$ are $v$-independent functions and $\omega$ is a $v$-independent one-form, all of them satisfying particular differential equations on the transverse three-dimensional space.

\item {\bf Time-like class} There exist local coordinates $t, x^{i}$, $i = 1, \hdots, 4$, such that the metric can be written as:
\begin{equation} \label{time-like_class}
ds^{2} =  -Z^{-2} (dt + \omega)^2 + Z^{-1} g_{ij} dx^{i} dx^{j}
\end{equation}

\noindent
where $Z$ is $t$-independent and $\omega$ is a $t$-independent one-form. The symbol $g_{ij}$ denotes a four-dimensional euclidean metric, which has to be Hyper-K\"ahler. In fact, it can be shown that the triplet $\Phi^{s}$ of two forms is the corresponding Hyper-K\"ahler structure. Therefore, time-like solutions are amenable to be locally written in terms of a four-dimensional Riemannian manifold, making the problem of obtaining these solutions a problem in Riemannian geometry and suggesting that the moduli-space problem of time-like solutions is well-defined.
 
\end{itemize}

\noindent
Once the local form of the solution has been found, in principle one can obtain the global solution by a maximally analytic extension of the space-time, which is a very non-trivial procedure that has to be done on a case by case basis. In the analytic extension, the fields of the solution can potentially change some of their properties which were holding on the original local set. 

As an example of this global phenomenon, one can analyze the \emph{chirality} of supersymmetry spinor $\epsilon$ as one moves on the space-time manifold. There is no notion of chirality in five dimensions, an it has to be imported from the four-dimensional language. Given the structure of the time-like class of supersymmetric solutions, amenable to be written in terms of a four-dimensional, \emph{transverse} space, together with the fact that the spinor does not depend on the time coordinate, one can study $\epsilon$ as a Clifford module in four euclidean dimensions. Notice that $\epsilon$ remains irreducible as a $Cl(4)$ module, but is not irreducible anymore as a representation of $Spin(4)$. The Clifford module is still of quaternionic type, and seeing $\epsilon$ as a Clifford-module for $Cl(4)$ it can be decomposed as follows:

\begin{equation}
\epsilon = \epsilon_{+} \oplus \epsilon_{-}
\end{equation}

\noindent
where $\epsilon_{\pm}$ are $Spin(4)$ spinors of positive and negative chirality, namely inequivalent irreps. of $Spin(4)$. Now, $\epsilon$ is parallel under a generalized connection on the spinor bundle $S$ given by the Killing spinor equation. In other words, the Killing spinor equation \eqref{eq:KSE5} can be rewritten as:

\begin{equation}
D_{v} \epsilon = 0 \, , \qquad \forall\, v\in \mathfrak{X}(M_{5})
\end{equation}

\noindent
where $D\colon \Gamma(S)\to\Gamma(S)\otimes \Omega^{1}(M_{5})$ is a connection on the spinor bundle. Therefore, if $\epsilon$ is non-zero at one point it will be non-zero everywhere. However, this does not imply that the individual components $\epsilon_{\pm}$ must remain non-zero at every point: they can indeed vanish at a locus in $M_5$, as long as they do not vanish simultaneously. This fact was observed in~\cite{Martelli:2003ki,Babalic:2014fua,Babalic:2014dea,Babalic:2015kka,Babalic:2015xia} for the case of M-theory compactifications on eight-manifolds. In these references it was proven that the fact that $\epsilon$ can become chiral at some loci  has crucial and remarkable consequences on the geometry of the internal space, which in turn implies that it is equipped with a particular singular foliation and a stratified $G$-structure. In particular, in~\cite{Grana:2014rpa,Babalic:2014fua,Babalic:2014dea,Babalic:2015kka,Babalic:2015xia} it was shown that the standard theory of $G$-structures is not enough to describe general String/M-Theory compactifications, but that a more general mathematical theory is needed. It seems that this more general theory is then the theory of stratified $G$-structures and foliations. 

On a similar vein, one can consider the global properties of the Killing vector field $p$, which concern the main analysis of this note. Supersymmetric solutions of five-dimensional, pure, $\mathcal{N}=1$ Supergravity are characterized in terms of a polyform $E$ satisfying some differential equations that can be succinctly written as follows:

\begin{equation}
D^{Ad}\, E = 0 
\end{equation}

\noindent
where $D^{Ad}$ is the connection on the K\"ahler-Atiyah bundle induced by the \emph{supersymmetric} connection~\cite{Lazaroiu:2012du}:

\begin{equation}
D_{u}\equiv \nabla_{u} - \frac{1}{8\sqrt{3}}\, u^{\flat}\wedge F + \frac{1}{2\sqrt{3}}\, \iota_{u}\, F  \qquad \forall u\in\mathfrak{X}(M_{5})
\end{equation}

\noindent
We then see that $E$ is parallel under the connection $D^{Ad}$, and therefore if it is non-zero at one point (which holds by assumption), it will be non-zero everywhere. However, this does not imply that the vector field $p$ or its norm is parallel under any connection. Therefore, it is in principle possible that the norm of $p$ becomes null at some locus in $M_{5}$.

Inspired then by these results, we want to explore if the following phenomenon may happen in five-dimensional, pure $\mathcal{N}=1$ Supergravity: we want to check if there are supersymmetric solutions of this theory having a Killing vector whose norm has an infinite number of vanishing derivatives at its null locus in $M_{5}$, and is generically time-like on its complement. In the next sections we explicitly construct a family of these solutions, which motivates a formal study of their global geometry and topology that will be presented elsewhere~\cite{Pinfoliations}.


\section{Five-dimensional $\mathcal{N}=1$ smooth solutions asymptotic to $AdS_3 \times S_2$}
\label{sec:model}

\subsection{Review}

 In this section we review the class of five-dimensional solutions in minimal Supergravity we will be concerned with. These are smooth $\mathcal{N}=1$ solutions that asymptote to $AdS_3\times S_2$ and have a Gibbons-Hawking (GH) space as four-dimensional Hyper-K\"{a}hler base.  As we will show at the end of this section, these solutions admit a Killing vector that is time-like almost everywhere, except for some codimension one loci where it becomes null. This fact is crucial for the construction described in the following sections that leads to a solution where the norm of the Killing vector vanishes at a point of the GH space together with an infinite number of derivatives.
  
The conditions that ensure the smoothness of the class of $\mathcal{N}=1$ five-dimensional solutions we review here were first found in~\cite{Bena:ring, Berglund:microstates}, while we follow the conventions of~\cite{deBoer:AdS_3, Bena:family} to impose the solutions to asymptote to $AdS_3 \times S_2$. As the Killing vector is time-like almost everywhere the metric can be written in the standard form of the time-like class~\eqref{time-like_class}:
\begin{align}
 ds^2_{5}& =-Z^{-2} (dt+\omega)^2+ Z \, g_{ij} dx^i dx^j \label{metric} \\
  A  &=  -Z^{-1}(dt+\omega) +V^{-1}K \, (d\psi + \omega_0) + \xi \label{potential}
\end{align}
where $A$ is the potential one-form and $g_{ij} dx^i dx^j$ is a Gibbons Hawking metric:
\begin{equation}\label{metric_gh}
g_{ij}dx^i dx^j=V^{-1} (d\psi+\omega_0)^2+V\delta_{ab} dx^adx^b
\end{equation}
with $a,b=1,2,3$ and the GH fiber $\psi$ has period $4\pi$. The functions in~\eqref{metric}, \eqref{potential} and~\eqref{metric_gh} are defined on the three-dimensional space spanned by $x=(x^1, x^2, x^3)$ in the GH space. In particular, $V$ in~\eqref{potential} and~\eqref{metric_gh} is a harmonic function
\begin{align}
V=\sum_{i=1}^N \frac{v_i}{r_i} \quad \quad \quad  r_i=|x-g_i|\label{function_V}
\end{align}
where $g_i$ is the location of the $i$-th pole (GH center). The one-form $\omega_0$ in~\eqref{metric_gh} is related to $V$ via $d\omega_0 =\star d V$, where the Hodge star is constructed using the euclidean $\mathbb{R}^3$ metric.  By imposing $v_i \in \mathbb{Z}$, the poles of $V$ become orbifold singularities for the metric~\eqref{metric_gh} (which are benign in String Theory) of the form $S_3/\mathbb{Z}^{|v_i|}$. One has to impose also  $\sum_i v_i =0$  so that the metric asymptotes to $AdS_3\times S_2$\footnote{In~\cite{Bena:ring, Berglund:microstates} these solutions are built to be microstates for a class of five-dimensional three-charge black holes, and hence they asymptote to flat space. This is achieved by requiring $\sum_i v_i$ =1.}.

The function $Z$ in~\eqref{metric} and~\eqref{potential} is expressed as a combination of $V$ together with two additional harmonic functions $K$ and $L$:
\begin{equation}\label{warpfactors}
 Z=L+\frac{K^2}{V}
 \end{equation}
Requiring $Z$ to be smooth everywhere constrains $K$ and $L$ to have the same poles as $V$ in~\eqref{function_V} and uniquely fixes the residues of $L$ once those of $K$ have been specified\footnote{In~\cite{Bena:ring, Berglund:microstates} it was necessary to add a constant $+1$ to $L$ so that $Z$ is nicely behaved at infinity for an asymptotically-flat metric. As our solutions are asymptotic to $AdS_3 \times S_2$ there is no such requirement for $Z$.}:
\begin{equation} \label{function_K}
K=\sum_{i=1}^N \frac{k_i}{r_i} \quad \quad \quad L = - \sum_{j=1}^N \frac{k_j^2}{v_j}\frac{1}{r_j} 
\end{equation}
The one-form $\xi$ in~\eqref{potential} is then defined by $ d\xi =-\star dK$, where the Hodge star is again defined using a flat $\mathbb{R}^3$ metric. 
The BPS solution for the angular momentum one-form $\omega$ in~\eqref{metric} is written as  
  \begin{equation}\label{momentum_k}
 \omega = S (d\psi +\omega_0)+\zeta
 \end{equation}  
 with $ S$ given by
  \begin{equation}\label{mu}
 S=\frac{K^3}{V^2}+\frac{3KL}{2V}+M
 \end{equation}  
 where $M$ is another harmonic function that has the same poles as $V$ in~\eqref{function_V}. Its residues are fixed by those of $V$ and $K$ so that $S$ is finite and smooth everywhere:
 \begin{equation}\label{function_M}
    M = m_0 +\frac{1}{2} \sum_{i=1}^N \frac{k_i^3}{v_i^2} \frac{1}{r_i}
 \end{equation}
The nonzero constant $m_0$ in~\eqref{function_M} determines the radius of the asymptotically $AdS_3\times S_2$ metric~\eqref{metric}. Finally, $\zeta$ in~\eqref{momentum_k} is given given by
  \begin{align}
 \star d\zeta = V dM - M dV +\frac{3}{2}(KdL-LdK) 
  \end{align}

Some constraints have to be satisfied to prevent the existence of closed time-like curves. First of all, from~\eqref{metric} it easy to see that one has to require $S$ in~\eqref{mu} to vanish at each GH center. This imposes $N-1$ independent conditions known as \textit{bubble equations} that involve the residues $v_i$ and $k_i$ together with the inter-center distances $r_{ij}$:
\begin{align} \label{bubble_equations}
\sum_{\substack{j=1  \\ j\neq i}}^N \left(\frac{k_j}{v_j} - \frac{k_i}{v_i}\right)^3 \frac{v_i v_j}{r_{ij}} + v_i \, m_0 &=0 &   i&=1,...,N-1
\end{align}
Secondly, to avoid closed time-like curves the following inequalities must hold everywhere in the GH base space:
\begin{equation} \label{ctc}
Z^3V-S^2 V^2 >0 \quad \quad \quad \quad ZV>0
\end{equation}
Therefore, to completely determine one of these solutions one has to fix $m_0$ in~\eqref{function_M} and the number of GH centers $N$. Secondly, one specifies the residues $v_i$, $k_i$ and the inter-center distances $r_{ij}$ so that~\eqref{bubble_equations} are satisfied. There is no general prescription known to satisfy~\eqref{ctc} and these two conditions have to be checked a posteriori.

It is useful to show that the metric~\eqref{metric} asymptotes to $AdS_3 \times S_2$. We parameterize the $\mathbb{R}^3$ base of the GH space with spherical coordinates $r, \theta, \phi$ and define the following quantities:
\begin{align}
Q &= \sum_{i=1}^N k_i   &    J&=\left| \sum_{i=1}^N v_i \cdotp g_i \right|         &         P&=\sum_{j=1}^N \frac{k_j^2}{v_j}   \label{electric_charge}
\end{align}
By introducing the coordinates~\cite{deBoer:AdS_3}:
\begin{align}
\eta &= Q \log \frac{r}{Q}   &   \tau&= t  &  \sigma&= 2 m_0 \psi -t 
\end{align}
the metric~\eqref{metric} and potential~\eqref{potential} asymptotically become
\begin{align}
ds^2 &\simeq d\eta^2+\,\frac{1}{4m_0}e^{\frac{\eta}{Q}}(-d\tau^2+d\sigma^2) + Q^2\left(d\theta^2 +\sin^2\theta (d\phi+\tilde{\omega}_0)^2\right) \label{metric_asymptotic} \\
A &\simeq Q \cos \theta \, (d \phi +\tilde{\omega}_0) + -\frac{3P}{2Q m_0} \, (d\sigma +d  \tau) \label{potential_asymptotic}
\end{align}
where we have defined
\begin{equation}
\tilde{\omega}_0=\frac{J}{2 Q^3}(d\tau - d\sigma)  
\end{equation}

Equation~\eqref{metric_asymptotic} shows that the metric of these solutions asymptotes to $AdS_3\times S_2$\footnote{The first part of the metric~\eqref{metric_asymptotic} is written as a Poincaré-AdS space, as the coordinate change to AdS becomes involved for solutions with more that two centers, where the GH metric coincides with a Eguchi-Hanson metric~\cite{Bena:family}.}. In addition, from~\eqref{metric_asymptotic} it is clear that one should impose $Q$ in~\eqref{electric_charge} to be positive.

Note that the asymptotic behavior of these solution~\eqref{metric_asymptotic} and~\eqref{potential_asymptotic} is uniquely determined once the quantities in~\eqref{electric_charge} have been fixed.



\subsection{From time-like to null Killing vector}

The $\mathcal{N}=1$ five-dimensional Supergravity solutions reviewed in Section~\ref{sec:model} have a Killing vector $p$ that is time-like almost everywhere and that in the coordinates of~\eqref{metric} is simply written as $\frac{\partial}{\partial t}$. As recalled in Section~\ref{sec:5dSugra}, it is locally defined  by the Killing spinor through a bilinear form $\mathcal{B}$, which for the metric~\eqref{metric} simply becomes:
\begin{equation} \label{killing_vector}
p_{a}=\mathcal{B}_0 (\epsilon, \gamma_{a} \epsilon) = \bar{\epsilon} \, \gamma _{a} \epsilon
\end{equation}
while~\eqref{ctc} allows to rewrite~\eqref{Fiertz} as follows:
\begin{equation}\label{killing_norm}
g_5(p,p)=-Z^{-2} = -(ZV)^{-2} \,  V^2
\end{equation}
 Therefore in our class of solutions the Killing vector is time-like everywhere except for the $V=0$ loci, where it becomes null. This peculiarity belongs to the class of solution reviewed in Section~\ref{sec:model} and the reason why the time-like Killing vector can become null at some loci lies in the fact that the GH metric~\eqref{metric_gh} is ambipolar. Indeed, its signature can pass from $(-,-,-,-)$ to $(+,+,+,+)$ and the surfaces $V=0$ mark the borders between the two signatures. This does not affect neither the physical validity of the Supergravity solutions of Section~\ref{sec:model} nor their smoothness, as it can be shown~\cite{Bena:lecture_notes} that the full metric~\eqref{metric} has lorentzian signature everywhere. 
 
 To better understand how these solutions can switch from the time-like class to the null one it is useful to analyze what happens to the Killing spinor $\epsilon$ as one approaches the $V=0$ loci~\cite{Bena:species}. The standard frames for the metric~\eqref{metric} are given by:
 \begin{align}
 e^0 &= Z^{-1} (dt + \omega) &  e^1&=(ZV)^{\frac{1}{2}} V^{-1} (d\psi +\omega_0) & e^{i+1} &= (ZV)^{\frac{1}{2}} dx^i  \label{frames}
 \end{align}
 and it was shown in~\cite{Gauntlett:2002nw} that if $e^0$ is written as in~\eqref{frames} then the Killing spinor satisfies
 \begin{equation} \label{spinor_time-like}
 \gamma^0 \epsilon = \epsilon
 \end{equation}
 in the frame indices defined by~\eqref{frames}. This is indeed the prototypical spinor equation for time-like solutions. However, the frames $e^0$ and $e^1$ in~\eqref{frames} become singular as $V$ approaches zero and~\eqref{spinor_time-like} does not hold in these regions. In particular, one can see from~\eqref{killing_norm} that
 \begin{equation}
 \epsilon^\dagger \epsilon = \bar{\epsilon} \gamma^0 \epsilon = Z^{-1} = (ZV)^{-1}V   \label{spinor_norm}
 \end{equation}
 which shows that the norm of $\epsilon$ vanishes in the $V=0$ loci. To understand what happens in these regions one has to define a completely regular set of frames, which is made possible by simply replacing $e^0$ and $e^1$ in~\eqref{frames} with:
 \begin{align}
 \hat{e}^0 &= \frac{1}{2}(V + V^{-1}) \, e^0 +\frac{1}{2}(V-V^{-1}) \, e^1 \nonumber \\
    \hat{e}^1 &= \frac{1}{2}(V - V^{-1}) \, e^0 +\frac{1}{2}(V+V^{-1}) \,  e^1 
  \end{align}
  Note that the regular set of frames is related to the original one~\eqref{frames} simply by a boost, with boost parameter $\chi$ defined by
  \begin{align}
  \cosh \chi &=\frac{1}{2}(|V|+|V|^{-1}) &   \sinh \chi &=\frac{1}{2} ( |V| - |V|^{-1})
  \end{align}
  and clearly the boost parameter becomes infinite when $V=0$. The Killing spinor in the regular set of frames $\hat{\epsilon}$ is hence related to the original one $\epsilon$ by
  \begin{equation}
  \hat{\epsilon} = e^{\frac{\chi}{2}\gamma^{01}} \epsilon = \frac{1}{2} |V|^{\frac{1}{2}} ( 1+ \gamma^{01}) \epsilon + \frac{1}{2} |V|^{-\frac{1}{2}}(1-\gamma^{01}) \epsilon \label{spinor_boost}
  \end{equation}
  where the frame indices are again defined by~\eqref{frames}. From~\eqref{spinor_norm} one can see that the magnitude of $\epsilon$ in the original frame vanishes as $|V^{\frac{1}{2}}|$  and then from~\eqref{spinor_boost} one concludes that on the $V=0$ loci the spinor $\epsilon$ satisfies
  \begin{equation}
  \gamma^{01} \epsilon = - \epsilon
  \end{equation}
  which is nothing but the prototypical spinor equation for the null class of five-dimensional solutions.

 As $V$ in~\eqref{function_V} is a harmonic function, these loci are two-dimensional surfaces in the GH space~\cite{Bena:species}, hence the Killing vector cannot become null on an open set of the five-dimensional space-time manifold. The closest condition to having a null Killing vector on an open set is to have it vanish at a point together with all its derivatives. This result can be achieved if one compromises to consider a solution of the kind reviewed in Section~\ref{sec:model} with an infinite number of GH centers. Indeed, in the next section we show that by suitably arranging a solution with $2N$ GH centers $V$ and all its derivatives up to order $2N-2$ can be set to zero at a point. In the limit where $N$ becomes infinite $V$ and all its derivatives vanish at a point of the GH space. Consequently, all derivatives of the norm of the Killing vector of the limit solution vanish at the origin. Since the null condition is a closed condition, we conclude that:
 
 \begin{itemize}
 	\item The norm of the Killing vector field is not a real analytic function at the origin.
 \end{itemize}


\section{Our construction}
\label{sec:construction}

In this section we show how to arrange a smooth solution of the class reviewed in Section~\ref{sec:model} with $2N$ GH centers so that all the derivatives of $V$ up to order $2N-2$ vanish at a point, which we fix to be the origin of the GH space.

We consider a solution with $2N$ GH centers located on the same axis, parameterized by the coordinate $x$ in the GH base space. The full solution~\eqref{metric} then has cylindrical symmetry with respect to this axis and the angular coordinate can be ignored. Focusing on a plane containing the axis, the orthogonal direction is parameterized with $y$. We dispose $N$ GH centers on the semi-axis $x\geq 0$, labeled by $i=1,...N$, while the remaining $N$ centers are arranged on $x<0$ and labeled by $i=-1,-2,...-N$. Denoting the distance of the $i-$th center from the origin with $d_i$, we constrain the residues $v_i$'s and $k_i$'s of~\eqref{function_V}, \eqref{function_K} and the $d_i$'s to satisfy the following conditions:
\begin{align}
v_{-i}=-v_i \quad \quad \quad k_{-i} = k_i \quad \quad \quad d_{-i}=d_{i} \quad \quad \quad  i=1,...N  \label{ansatz_1}
\end{align}
In addition, for the semi-axis $x>0$ we require the sign of the $v_i$'s to be alternating: 
\begin{equation}\label{ansatz_2}
v_{i}=(-1)^{i+1} |v_i| \quad  \quad \quad  i=1,...N 
\end{equation}
As a consequence of~\eqref{ansatz_1} the quantity $P$ defined in~\eqref{electric_charge} is identically zero. In addition, because of~\eqref{ansatz_1} and~\eqref{ansatz_2}, only $N$ equations~\eqref{bubble_equations} among the initial $2N-1$ remain independent, as the equation for $i$ is equivalent to that for $-i$. 

This configuration ensures that $V$ in~\eqref{function_V} and some of its derivatives vanish at the origin $x=y=0$. In particular, by introducing the notation
\begin{equation}
(m,n)\equiv \left. \frac{\partial ^{m+n} V}{\partial x^m \partial y^n}\right \rvert_{x=y=0}
\end{equation}
one can observe the following simplifications:
\begin{itemize}
\item $(2s, n) =0 $ $\forall s, n$, 
\item $(2s+1, 2p+1) =0$ $\forall s,p$,
\item $(m,n)$ and $(s,p)$ are multiples of each other provided that $m+n = s+p$.
\end{itemize}
Therefore, after fixing the number of GH centers to be $2N$ and requiring~\eqref{ansatz_1} and~\eqref{ansatz_2} to hold, one can annihilate the $N-1$ nonzero derivatives at the origin of the form $(1, 2s)$ with $s=0,...N-2$. Indeed, defining the ratios $\xi_i = v_i/v_1$ and $\delta_i = d_i/d_1$, these $N-1$ constraints completely fix the $N-1$ $\xi_i$'s as functions of the $\delta_i$'s, so that all the derivatives up to order $2N-2$ are zero at the origin. For instance, in a solution with $2N=8$ GH centers one is free to assign the three ratios of the distances $\delta_i$'s and then the three $\xi_i$'s are determined to annihilate the derivatives $(1,0)$, $(1,2)$ and $(1,4)$. As a result, $V$ and all its derivatives up to order six are zero at the origin.

Without loss of generality we fix
\begin{equation}\label{convention}
v_1 =1 \quad \quad \quad \quad d_1 =1
\end{equation}
so that the $v_i$'s can be directly expressed as functions of the $d_i$'s. There are two such expressions, depending on whether the GH centers $g_i$ with $|i|>1$ are added externally with respect to $g_1$ and $g_{-1}$, namely choosing $d_i>d_j$  for $i>j$, or internally, satisfying $d_j<d_i$ for $j>i$. We adopt this second convention, as the numerical investigations for the limit $N\rightarrow \infty$ give evidence that this is the physically sensible option, presenting the formula we got from the first option in Appendix~\ref{appendix:external}. 

Therefore, after fixing $N$, one arranges the first two centers $g_{\pm 1}$ on the line by means of~\eqref{convention}. Then the remaining centers are progressively added in the intervals $]0,1[$ and $]-1,0[$ according to~\eqref{ansatz_1} and~\eqref{ansatz_2}. We are left free to assign the distances subject to the requirement $d_i < d_j <1$ for $i>j$, while the $v_i$'s remain fixed to annihilate all the derivatives of $V$ up to order $2N-2$ at the origin:
\begin{equation} \label{internal_vanishing}
|v_i| = d_i^{2N-2}\prod_{\substack{j=2  \\ j\neq i}}^N  \frac{1-d_j^2}{|d_i^2-d_j^2|} \quad \quad \quad i=1,...N
\end{equation}
and their sign is determined according to~\eqref{ansatz_2}.

To completely determine a five-dimensional solution of the class reviewed in Section~\ref{sec:model}, also $N$ parameters $k_{i}$'s for $i=1,...N$ have to be determined. These parameters are fixed by numerically solving the $N$ remaining independent equations~\eqref{bubble_equations} while requiring $Q>0$ in~\eqref{electric_charge}. 

To summarize, after fixing $N$, one disposes the centers symmetrically with respect to the origin according to~\eqref{ansatz_1} and ~\eqref{ansatz_2} and fixes the $N$ distances $d_i$'s subject to $0<d_i<d_j<1$ for $i>j$.  The $v_i$'s are determined by~\eqref{internal_vanishing} and the $k_i$'s are determined by solving the $N$ independent bubble equations~\eqref{bubble_equations}. Once the $v_i$'s and $k_i$'s have been found, the whole five-dimensional solution can be written according to Section~\ref{sec:model}. 

The $N$ bubble equations have to be solved numerically once $m_)$ and the distances $d_i$'s have been fixed. We studied the behavior of these solutions for many different distance distributions $d_i$, progressively increasing $N$. Our numerical analysis indicates that among the many possible real solutions for the $k_i$'s, there is always one where the $k_i$'s are all positive. We observed that this is the only option to systematically satisfy $VZ>0$ everywhere in~\eqref{ctc}, as we verified numerically that if some of the $k_i$'s are negative the function $VZ$ can become negative close to some of the GH centers. This kind of analysis was repeated for many different distance distributions $d_i$ using~\eqref{internal_vanishing} and also adding the GH centers externally with respect to $g_{\pm 1}$ using~\eqref{external} in Appendix~\ref{appendix:external}.  As changing the value of $m_0$ in~\eqref{bubble_equations} just scales the values for the $k_i$'s, we fix $m_0=-1$ in our numerical analysis, keeping in mind that our results are valid for any nonzero $m_0$.

We found that the second constraint in~\eqref{ctc} is also satisfied if one requires the $k_i$'s to be positive for different distributions of distances and different $N$, although we lack a rigorous generalization to arbitrary $N$.


\section{Analysis of the solutions in the limit of infinite centers} 
\label{sec:limit}
In this section we analyze the limit $N\rightarrow \infty$ applied to the construction of Section~\ref{sec:construction}. In this limit all the derivatives of $V$ vanish at the origin. For finite $N$ the construction of Section~\ref{sec:construction} leads to valid, smooth Supergravity solutions of the class reviewed in Section~\ref{sec:model}. However, the situation is different for $N\rightarrow \infty$, as one gets a solution with an infinite numbers of poles. Depending on the chosen distribution for the $d_i$'s, the GH centers can accumulate towards single points or in finite intervals, but one cannot say a priori whether this invalidates the smoothness of the solution in the limit without a thorough analysis of the metric~\eqref{metric} in the limit $N\rightarrow \infty$.

Two more general issues can compromise the physical relevance of the limit solution for $N\rightarrow \infty$. First of all, the constraint $Z^3V-S^2V^2>0$ should be checked for every $N$. We verified that for a wide range of distributions of the $d_i$'s and for different values of $N$ this condition is valid and it seems reasonable to assume that it still holds in the infinite limit. Secondly, it is natural to require that the limit solution still asymptotes to $AdS_3\times S_2$, and hence that its metric and potential can be rewritten as in~\eqref{metric_asymptotic} and~\eqref{potential_asymptotic} at infinity. In particular, this means that $J$ and $Q$ in~\eqref{electric_charge} have to remain finite for $N\rightarrow \infty$\footnote{We remind the reader that because of~\eqref{ansatz_1} $P$ in~\eqref{electric_charge} is identically zero in our construction.}. As the sign of the $v_i$'s is alternating according to~\eqref{ansatz_1} and~\eqref{ansatz_2} and as we are accumulating the poles within finite intervals, it is easy to find a distance distribution for the poles such that $J$ in~\eqref{electric_charge} goes to zero in the limit.

On the other hand, it is tricky to find a distribution for the $d_i$'s so that the parameter $Q$ in~\eqref{electric_charge} does not grow indefinitely with $N$. As remarked at the end of Section~\ref{sec:construction}, for each finite $N$ the $k_i$'s are determined by numerically solving $N$ bubble equations~\eqref{bubble_equations} and selecting the only solution where they all have the same sign. For this reason $Q$ in~\eqref{electric_charge} can easily grow to become infinite with growing $N$. 

For $N$ large enough there seems to be a correlation between how the $v_i$'s -determined via~\eqref{internal_vanishing}- and the $k_i$'s behave with $N$. We observed that an exponential growth for the $v_i$'s is paralleled by an exponential growth of $Q$ in~\eqref{electric_charge}, which we want to avoid. 

We found a distribution for the $d_i$'s so that the $v_i$'s attain a finite limit for $N\rightarrow \infty$ that can lead to a finite $Q$ as well. For fixed $N$ we assign the distances $d_i$'s according to:
\begin{equation} \label{distribution_squashing}
d_i = 1-\frac{(i-1)^\alpha}{N^\beta}\quad \quad \quad i=1,...N
\end{equation}
with $0< \alpha < \beta$. In Appendix~\ref{appendix:convergence} we show that for each fixed $i$, $|v_i|$ given by~\eqref{internal_vanishing} with~\eqref{distribution_squashing} is finite in the limit provided that $\beta >1$. Note that~\eqref{distribution_squashing} satisfies $0<d_i < d_j\leq 1$ for $i>j$ and that in the limit $N\rightarrow \infty$ all the GH centers collapse on the fixed $g_1$ and $g_{-1}$ at unitary distance from the origin. This does not automatically imply that the limit solution become singular without a thorough analysis of the full metric~\eqref{metric} in the limit.

\begin{figure}[h]
\begin{center}
 \includegraphics[height=2.5cm]{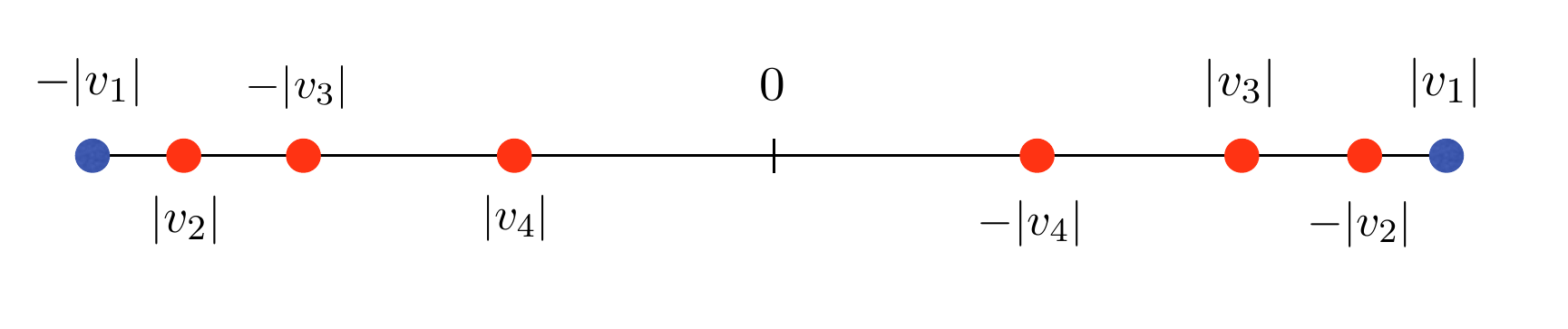}
 \end{center}
\caption{The configuration of GH centers obtained with the distance distribution~\eqref{distribution_squashing}. The blue dots represent the centers $g_{\pm 1}$, whose position has been fixed in~\eqref{convention}. The red dots represent the other GH centers, whose positions and $v_i$'s are determined by~\eqref{ansatz_1} and~\eqref{ansatz_2}.}
\label{figure_line}
 \end{figure}

It is important to stress that with the distribution~\eqref{distribution_squashing} the GH centers do not accumulate at the origin and hence the function $V$ in~\eqref{function_V} becomes constant and in fact it has an infinite number of vanishing derivatives in the limit. The configuration for the GH centers obtained with a distance distribution of the kind~\eqref{distribution_squashing} is represented in Figure~\ref{figure_line}.

\begin{figure}[h]
\begin{center}
 \includegraphics[height=8cm]{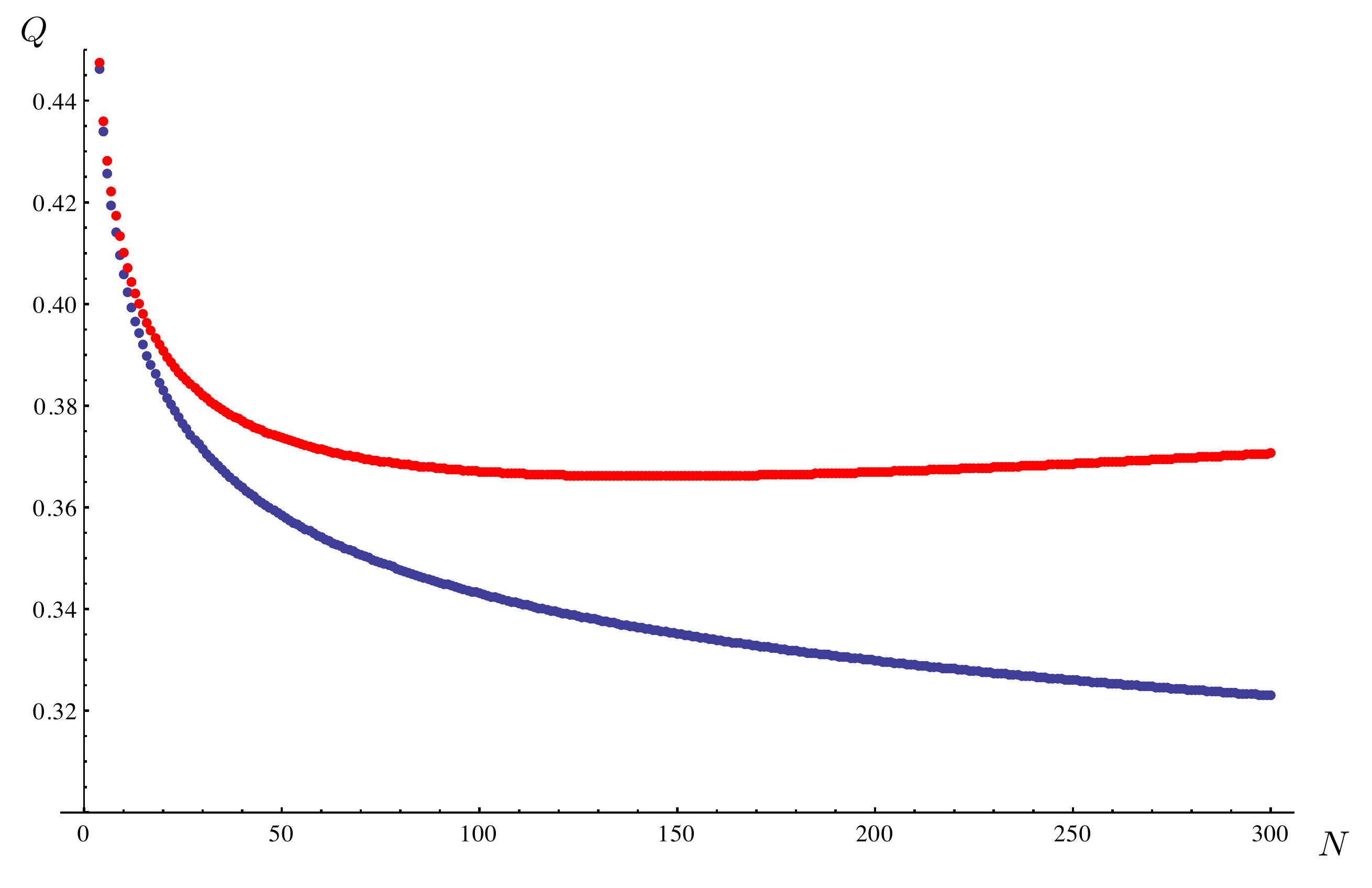}
 \end{center}
\caption{Numerical analysis for $Q(N)$ up to $N=300$ for two different regimes. The red dots were obtained choosing $\beta=2.05$ and $\alpha=1.99$ in~\eqref{distribution_squashing} and $Q(N)$ is divergent. The blue dots were obtained for $\beta=2.05$ and $\alpha=2$ and $Q(N)$ asymptotes to zero.}
\label{figure_charge}
 \end{figure}

We analyzed the behavior of $Q$ and $J$ in~\eqref{electric_charge} as a function of $N$ for different values of $\alpha$ and $\beta$ in~\eqref{distribution_squashing}, up to $N=300$\footnote{Note the solution obtained for $N=300$ has 600 GH centers in total. The related value of $Q$ is hence determined by numerically solving 300 bubble equations~\eqref{bubble_equations}, each of them consisting of 600 terms.}. The quantity $J$ in~\eqref{electric_charge} rapidly goes to zero as $N$ increases, while the behavior of $Q$ is more subtle. We noted that for $\beta-\alpha >1$ in~\eqref{distribution_squashing} the bigger $\beta -\alpha$ is the faster $Q(N)$ goes to zero for $N\rightarrow \infty$. For $\beta-\alpha<1$ we have not found a clear connection between this quantity and the behavior of $Q(N)$ at infinity. For instance, $Q(N)$ goes to infinity for $\beta=2.05$ and $\alpha=1.9$, while it goes to zero for $\beta=2.05$ and $\alpha=2$ - see Figure~\ref{figure_charge}. This indicates that there exists some value  $\tilde{\alpha}$ with $1.99<\tilde{\alpha}<2$ such that $Q(N)$ asymptotes to a constant. The related solution hence asymptotes to $AdS_3 \times S_2$ and the norm of the Killing vector has an infinite number of vanishing derivatives at the origin.

The asymptotic behavior of $Q(N)$ for $150\leq N \leq 300$ for the two distance distributions studied in Figure~\ref{figure_charge} is modeled by
\begin{equation}\label{interpolation}
Q(N) \sim \frac{N^a}{c\,N^b +d}
\end{equation}
where the parameters $a,b,c,d$ can be estimated numerically. The fact that $Q(N)$ for large $N$ behaves as in~\eqref{interpolation} is a general feature of the distance distribution~\eqref{distribution_squashing}. In particular, for the values of $\alpha$ and $\beta$ of Figure~\ref{figure_charge} we find the following values for $a,b,c,d$ in~\eqref{interpolation}:
 \begin{center}
\begin{tabular} { l  |c  c  c   c c }

                             &     $a$     &        $b$     & $c$        &         $d$    & $a-b$ \\
\hline
$ \alpha= 1.99$     &   1.83     &       1.79   &     3.51    &       -724.011 &  0.04 \\
$\alpha=2$          &    1.59     &      1.64    &   2.35      &     -44.12     &   -0.05 \\

\end{tabular}
\end{center}

\begin{figure}[h!]
\begin{center}
 \includegraphics[height=8cm]{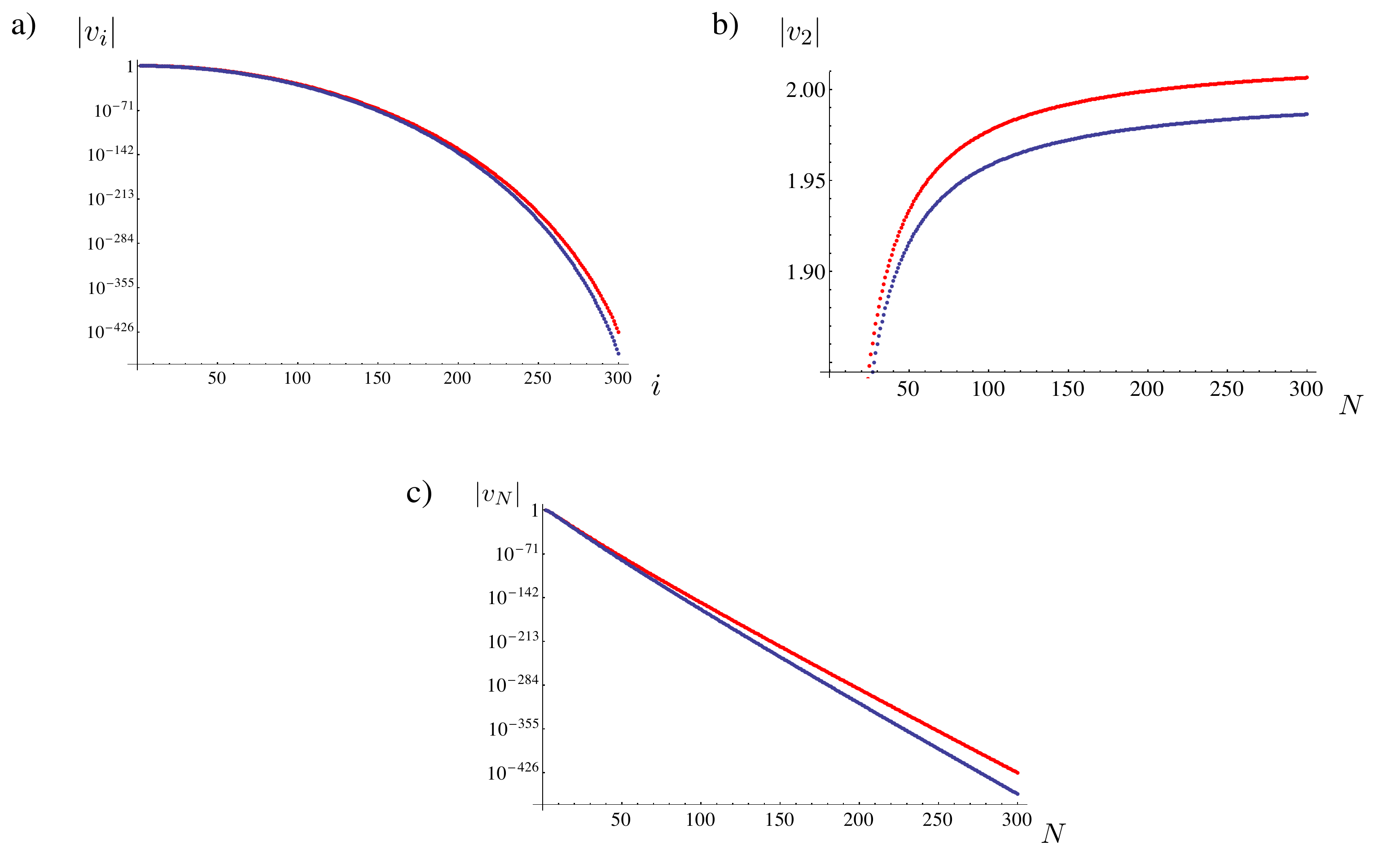}
 \end{center}
\caption{Numerical analysis for the $v_i$'s determined with~\eqref{distribution_squashing} and~\eqref{internal_vanishing} for $\alpha=1.99$, $\beta=2.05$ (red) and for $\alpha=2$, $\beta=2.05$ (blue). \textbf{a)} Representation of the $v_i$'s for the solutions with $N=300$. \textbf{b)} Trend of $|v_2|$ as a function of $N$ up to $N=300$. The trends of all the other $v_i$'s for fixed $i$ are similar. \textbf{c)} Trend of $|v_N|$ for the last-added GH center, as a function of $N$ up to $N=300$.}
\label{figure_tables}
 \end{figure}

It is interesting to study how the $v_i$'s determined by~\eqref{internal_vanishing} with the distance distribution~\eqref{distribution_squashing} evolve with $N$. From Figure~\ref{figure_tables} we see that for fixed $N$ the $v_i$'s of the last GH centers are small compared to the $v_i$'s to the first one. In addition, each $v_i$ slowly grows with $N$ but remains finite in the limit, as proved in Appendix~\ref{appendix:convergence} for the distance distribution~\eqref{distribution_squashing}. 

\begin{figure}[h]
\begin{center}
 \includegraphics[height=8cm]{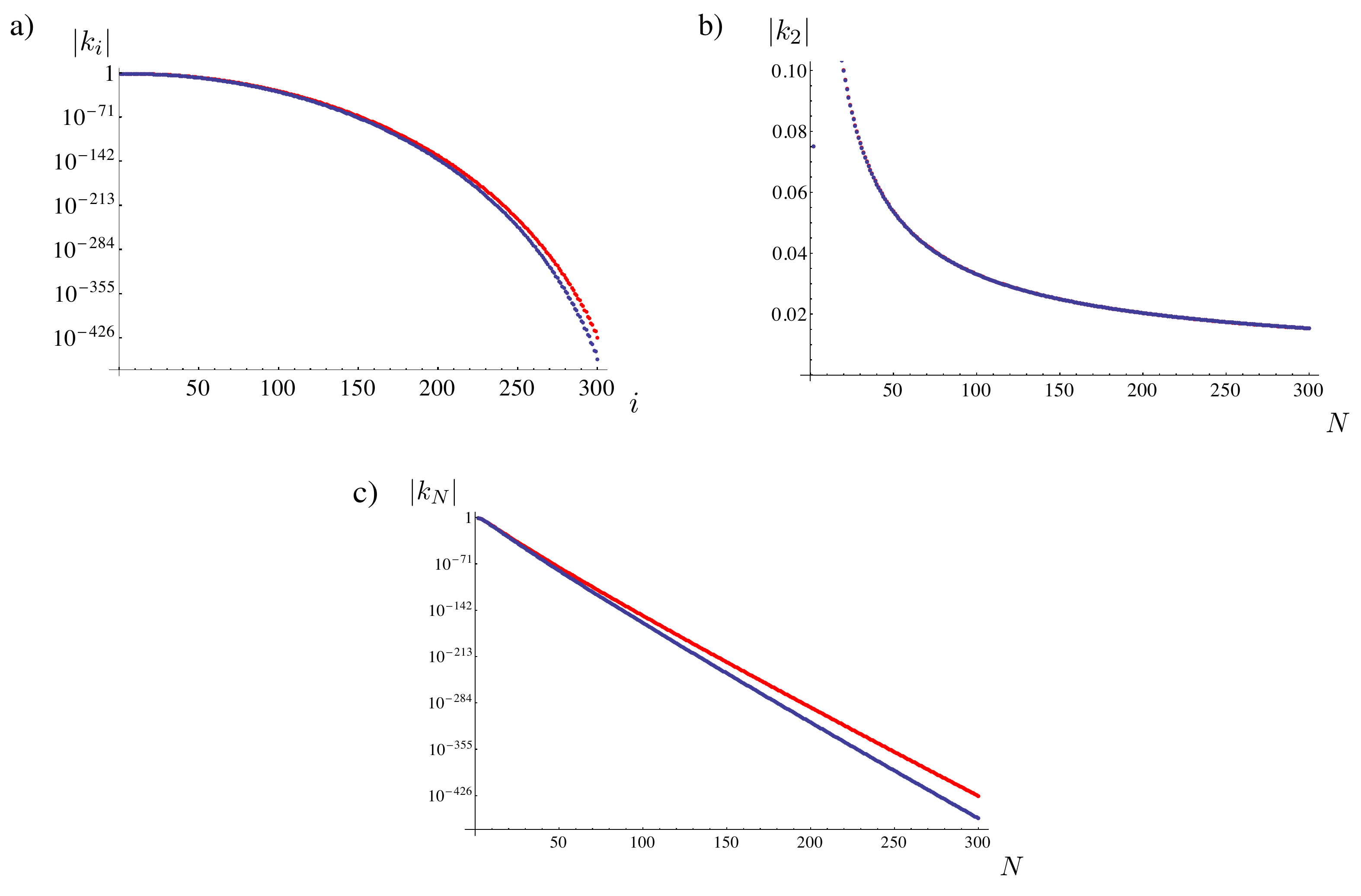}
 \end{center}
\caption{Numerical analysis for the $k_i$'s determined by solving the bubble equations~\eqref{bubble_equations} computed with the distance distribution~\eqref{distribution_squashing} for $\alpha=1.99$, $\beta=2.05$ (red) and for $\alpha=2$, $\beta=2.05$ (blue). \textbf{a)} Representation of the $k_i$'s for the solutions with $N=300$. \textbf{b)} Trend of $|k_2|$ as a function of $N$ up to $N=300$. The trends of all the other $k_i$'s for fixed $i$ are similar. \textbf{c)} Trend of $k_N$ for the last-added GH center, as a function of $N$ up to $N=300$.}
\label{figure_ktables}
 \end{figure}

A similar analysis can be performed to study the trend of the $k_i$'s, numerically determined by solving the bubble equations~\eqref{bubble_equations} - see Figure~\ref{figure_ktables}.
For a given $N$ the $k_i$'s of the last centers -namely the ones closest to the origin- are small compared to those of the most external centers. Each $k_i$ decreases with $N$ and limits to zero for $N\rightarrow \infty$ so that the conserved charge $Q$ in~\eqref{electric_charge} can remain finite in the limit. 

\newpage

\section{Conclusions and outlook}
\label{sec:conclusions}
We have shown the existence of five-dimensional Supergravity solutions such that the Killing vector together with an infinite number of derivatives vanish at a point in the base space. Since the null condition is a closed condition, this implies that the norm of the Killing vector is not a real analytic function at this point. This was achieved by considering the infinite limit for the number of centers in a class of smooth solutions with a Gibbons-Hawking base that asymptote to $AdS_3\times S_2$. By suitably tuning the distances of the poles before taking the limit, we gave evidence that the limit solution still asymptotes to $AdS_3 \times S_2$ and that its charges are finite. 

It is important to stress that while our construction is valid in Supergravity, it is not necessarily so in String Theory. Indeed, in Section~\ref{sec:construction} we stressed that the residues of $V$ should be integers, so that the Gibbons-Hawking space looks like an $S^3/\mathbb{Z}^{|v_i|}$ close to a center. In addition, also the residues of $K$ in~\eqref{function_K} should be constrained to be integers. This is because one can define numerous two-cycles by fibering the coordinate $\psi$ in~\eqref{metric_gh} between two centers and the differential of  $ A+Z^{-1}(dt+\omega)$ in~\eqref{potential} measures fluxes on these cycles that depend on the ratios $k_i/v_i$. The usual quantization conditions require these fluxes to be semi-integers. In the infinite limit analyzed in Section~\ref{sec:limit} one a priori cannot say that the $v_i$'s and the $k_i$'s are integers, although this does not seem impossible. It would be interesting to study whether there exist other distance distributions different from~\eqref{distribution_squashing} so that also this requirement is satisfied. 

It is also useful to analyze what happens in this class of solutions when the GH centers collide, as for the infinite limit with the distance distribution~\eqref{distribution_squashing}. In~\cite{Bena:lecture_notes} a physical interpretation for this phenomenon was given seeing these five-dimensional solutions as black hole microstates. Indeed, one can compute the metric distance between the most external GH center and a suitable cutoff far away in the Gibbons-Hawking space, which can be seen as the length of the would-be black hole throat in this language. This quantity is always finite for the solutions of Section~\ref{sec:model}, but can grow indefinitely once one moves the GH centers close together. As in the limit solution of Section~\ref{sec:limit} all the centers collapse on two points, this fact can be interpreted as the would-be black hole throat becoming infinite. 

Finally, the collapse of all the centers to two points does not automatically give rise to singularities in the metric. Indeed it was shown in~\cite{Bena:mergers, Bena:abyss} that what appears to be an essential singularity from the point of view of the three-dimensional base of the GH space can in fact give rise to a smooth solution once the full backreaction is taken into account. 
To verify the validity of this statement for our solutions it would be necessary to analyze the behavior of the function $Z$ in~\eqref{metric} close to $g_{\pm 1}$ in the infinite limit. This operation is highly nontrivial and we suspect that it might also not be well defined, as the limit for $N\rightarrow \infty$ might not commute with the limit $r\rightarrow g_{\pm 1}$. We plan to carry a full mathematical analysis of the properties of the limit solution in future work. 

\section*{Acknowledgments}
We are grateful to Iosif Bena for stimulating discussions. C.S.S. would like to thank Jos'e Figueroa - O'Farrill, Calin Lazaroiu and Tom\'as Ort\'in for very interesting and useful comments. The work of G.P. is supported in part by the John Templeton Foundation Grant 48222. The work of C.S.S. is supported in part by the ERC Starting Grant 259133 -- ObservableString.


\appendix

\section{Proof of convergence of the $v_i$'s for $N\rightarrow \infty$}
\label{appendix:convergence}

In this section we prove that each $v_i$ determined with~\eqref{internal_vanishing} with distance distribution $d_i$ given by~\eqref{distribution_squashing} with $0< \alpha < \beta$ is finite in the $N\rightarrow \infty$ limit, namely that
\begin{align}
\lim_{N\rightarrow \infty} |v_i| &= \lim_{N\rightarrow \infty} d_i^{2N-2}\prod_{\substack{j=2  \\ j\neq i}}^N  \frac{1-d_j^2}{|d_i^2-d_j^2|}  & \textmd{with} \quad \quad \quad   d_s&=1-\frac{(s-1)^\alpha}{N^\beta} \label{initial_limit}
\end{align}
is finite for each fixed $i$. 

As the logarithm is a monotonic bijection between $\mathbb{R}_+$ and $\mathbb{R}$, the $|v_i|$'s are finite if and only if $\log |v_i|$ remains finite, namely if
\begin{align}
\lim_{N\rightarrow \infty} \log |v_i| = \lim_{N\rightarrow \infty}  (2N-2) \log d_i + \sum_{\substack{j=2  \\ j\neq i}}^N \log \left( \frac{1-d_j^2}{|d_i^2-d_j^2|} \right) \label{limit_log}
\end{align}
 is finite. The first term on the rhs of~\eqref{limit_log} for sufficiently large $N$ becomes:
\begin{align}
(2N-2) \log d_i = (2N-2) \log \left(1-\frac{(i-1)^\alpha}{N^\beta} \right) \sim -2(i-1)^{\alpha} N^{1-\beta} 
\end{align}
and hence goes to zero provided that $\beta >1$. The sum in~\eqref{limit_log} for large $N$ can be rewritten as:
\begin{align}
\sum_{\substack{j=2  \\ j\neq i}}^N \log(j-1)^\alpha -\log |(j-1)^\alpha-(i-1)^\alpha| = - \sum_{\substack{j=2  \\ j\neq i}}^N \log \left| 1-\frac{(i-1)^\alpha}{(j-1)^\alpha}\right|
\end{align}
and then for $j>>i$ the asymptotic part of this sum becomes
\begin{align}
(i-1)^\alpha \sum_{j>>i}^N (j-1)^{-\alpha} \sim \frac{(i-1)^{\alpha}}{\alpha} N^{-\alpha -1} + const
\end{align}
which is finite provided that $\alpha > -1$. Therefore the limit~\eqref{initial_limit} with $0<\alpha <\beta$ is finite for each fixed $i$ provided that $\beta >1$.

\section{An alternative construction}
\label{appendix:external}

The distance distribution~\eqref{distribution_squashing} is the only one we have found that gives rise to a physically sensible solution in the limit $N\rightarrow \infty$ for the construction of Section~\ref{sec:construction}. As mentioned in Section~\ref{sec:construction}, after fixing the centers $g_{\pm 1}$ with~\eqref{convention} one can add the other centers externally with respect to these two, namely choosing $d_i >d_j>1$ for $i>j>1$. Then arranging the $2N$ centers on a line imposing~\eqref{ansatz_1} and~\eqref{ansatz_2} and requiring the derivatives of $V$ up to order $2N-2$ to vanish at the origin uniquely fixes the $v_i$'s as functions of the $d_i$'s:
\begin{equation}\label{external}
|v_i| =  (1+d_i)^{2N-2}\prod_{\substack{j=1  \\ j\neq i}}^N  \frac{d_j(2+d_j)}{| d_i(2+d_i)-d_j(2+d_j)|}
\end{equation}
and the sign of each $v_i$ is determined according to~\eqref{ansatz_2}.
We studied the solution in the limit for $N\rightarrow\infty$ using different distance distributions $d_i$'s. For instance, one can arrange the centers to be equally spaced on the axis, to accumulate close to a point or within an interval or to reach infinite distance with different spacings. We always find that $Q\rightarrow \infty$ for $N\rightarrow \infty$. The same happens inserting the GH centers internally using~\eqref{internal_vanishing} with distance distributions different from~\eqref{distribution_squashing}. We believe that the reason for this lies on the fact that each $|v_i|$ grows exponentially when one adds more and more centers. The distance distribution~\eqref{distribution_squashing} with~\eqref{internal_vanishing} is the only we have found that keeps the $v_i$'s finite in the limit, as shown in Figure~\ref{figure_tables}.


\renewcommand{\leftmark}{\MakeUppercase{Bibliography}}
\phantomsection
\bibliographystyle{JHEP}

\bibliography{References}


\label{biblio}

\end{document}